Research paper

# Falsification-driven reinforcement learning for maritime motion planning

Marlon Müller [a,b,c,*,1], Florian Finkeldei [a,*,1], Hanna Krasowski [b,*,1], Murat Arcak [b], Matthias Althoff [a,c]

[a] Technical University of Munich, Germany
[b] University of California, Berkeley, USA
[c] Munich Center for Machine Learning (MCML), Germany

ABSTRACT

Compliance with maritime traffic rules is essential for the safe operation of autonomous vessels, yet training reinforcement learning (RL) agents to adhere to them is challenging. The behavior of RL agents is shaped by the training scenarios they encounter, but creating scenarios that capture the complexity of maritime navigation is non-trivial, and real-world data alone is insufficient. To address this, we propose a falsification-driven RL approach that generates adversarial training scenarios in which the vessel under test violates maritime traffic rules, which are expressed as signal temporal logic specifications. Our experiments on open-sea navigation with two vessels demonstrate that the proposed approach provides more relevant training scenarios and achieves more consistent rule compliance.

## 1. Introduction

Safe navigation at sea requires vessels to comply with maritime traffic rules. These rules are essential for avoiding collisions, yet their situational and temporal dependencies are difficult to translate into control policies for autonomous vessels. Therefore, training RL controllers that achieve both rule compliance and task success is a central challenge for maritime autonomy. The design of the training environment and the selection of training scenarios are crucial when RL agents must learn complex behaviors. However, generating scenarios that reliably foster rule-compliant behavior is challenging. Falsification methods aim to provoke specification-violating behaviors in a system under test and typically provide counterexamples that demonstrate where a specification has been violated. While classical controllers are difficult to adapt directly based on such counterexamples, they can be incorporated into the RL training process to inform the agent about remaining failure modes and guide its decision-making toward improved specification compliance. In this article, we use counterexamples generated through falsification to improve the rule compliance of RL-controlled vessels. Specifically, we develop an efficient, optimization-based counterexample generation algorithm, which targets violations of maritime traffic rules expressed as signal temporal logic (STL) specifications by the current RL policy. Our experiments demonstrate that integrating falsification into the training process leads to improved rule compliance. Our main contributions are:

- We propose an efficient and practical falsification-driven RL algorithm that generates counterexamples for complex temporal logic traffic rules and uses them as training scenarios;
- We extend the maritime traffic rule formalization with robustness measures that are efficiently evaluable;
- We evaluate our approach using multiple falsification methods, demonstrating that falsification-driven RL improves traffic rule compliance.

The remainder of the article is structured as follows: We review related literature in Section 2 and introduce preliminary concepts in Section 3. The falsification-driven RL approach is detailed in Section 4, its application to maritime navigation is presented in Section 5, and evaluated in Section 6. We discuss the results in Section 7 and conclude in Section 8.

## 2. Related work

Our approach combines falsification with RL for maritime motion planning. Thus, we review related work on falsification and its combination with learning, advances in RL for maritime motion planning, and scenario generation for testing maritime systems.

* Corresponding author at: Technical University of Munich, Germany, and University of California, Berkeley, USA.
*E-mail addresses:* marlon.mueller@tum.de (M. Müller), florian.finkeldei@tum.de (F. Finkeldei), krasowski@berkeley.edu (H. Krasowski), arcak@eecs.berkeley.edu (M. Arcak), althoff@in.tum.de (M. Althoff).
[1] Equal contribution.

### 2.1. Falsification methods

The task of falsification is typically posed as an optimization problem that finds inputs to the environment such that the system under test violates a specification (Corso et al., 2022, Sec. 3.2). Several numerical optimization strategies have been successfully applied for falsification (Corso et al., 2022, Sec. 4), such as simulated annealing (SA) (Abbas et al., 2013; Yaghoubi and Fainekos, 2019), the cross-entropy method (Sankaranarayanan and Fainekos, 2012), evolutionary algorithms (Zou et al., 2014; Zhang et al., 2019; Waga, 2020; Sato et al., 2021), Bayesian optimization (Deshmukh et al., 2017; Shahrooei et al., 2023; Ramezani et al., 2025), and ant colony optimization (Annapureddy and Fainekos, 2010). A benefit of these numerical optimization strategies is that they can falsify systems with unknown dynamics, so-called black-box systems. In some cases, black-box falsification techniques are preferable, even for white-box systems, e.g., if the system dynamic is too complex (Corso et al., 2022, Sec. 2.2).

Besides numerical optimization, path planning algorithms can be used for falsification. These approaches construct an input sequence from an initial state to a set of failure states (Corso et al., 2022, Sec. 5), often using rapidly exploring random trees (Dreossi et al., 2015; Tuncali and Fainekos, 2019; Koschi et al., 2019). Path planning algorithms tend to perform well in high-dimensional state spaces. However, their performance deteriorates if the set of failures is small compared to the state space (Kim et al., 2006), and they require information about the current system state.

Lastly, RL-based falsifiers learn a policy that chooses the next system input based on its current state (Corso et al., 2022, Sec. 6). Their reward function is designed such that it reinforces falsifying actions (Yamagata et al., 2021). RL-based falsification is well-suited for searching for falsifying inputs over long time horizons, as only a policy is learned. However, the training is tedious and must be conducted for each system individually. In this article, we compare the numerical optimization methods covariance matrix adaptation evolution strategy (CMA-ES) (Hansen, 2016), particle swarm optimization (PSO) (Kennedy and Eberhart, 1995), and SA (Rere et al., 2015) to create falsifying scenarios as these methods are suitable for higher-dimensional nonlinear system dynamics (Zhang et al., 2019; Khandait et al., 2024).

### 2.2. Falsification for reinforcement learning

The concept of improving the performance of RL controllers through adversaries is used in Pinto et al. (2017), which alternately trains an RL adversary alongside the RL protagonist. The reward function of the adversary incentivizes failures of the protagonist to obtain more robust policies. The work in Pan et al. (2019) extends (Pinto et al., 2017) to turn-based interactions and augments the action-value function with a risk term based on the variance of its outputs across different models. Instead of training two competing RL agents, the work in Wang et al. (2020) generates counterexamples using the cross-entropy method to improve the training of RL agents. Its application for a braking assistance and an adaptive cruise control system shows improved generalization and safety of the resulting RL controller. A similar approach is Gangopadhyay and Dasgupta (2021), which uses Bayesian optimization as a different falsification strategy, applied to multi-agent RL tasks. Similar to Wang et al. (2020), we add falsifying scenarios to the RL scenario set during training, while considering more complex STL specifications as falsification targets.

### 2.3. Reinforcement learning for maritime navigation

RL approaches for maritime navigation have been increasingly proposed with the success of model-free RL. The investigated applications range from maneuvering in harbors (Zweigel et al., 2019) to navigating in rivers (Wang et al., 2019; Enevoldsen et al., 2023) to open-sea navigation (Johansen et al., 2016; Kufoalor et al., 2019; Krasowski and Althoff, 2024). Among model-free RL algorithms, Proximal Policy Optimization (PPO) (Schulman et al., 2017) has shown strong performance (Larsen et al., 2021; Cui et al., 2023) and is used in many RL approaches that inform the agent about maritime traffic rules through the reward (Zhao and Roh, 2019; Rongcai et al., 2023; Y. Wang et al., 2024). The simulated traffic for these RL agents is either non-reactive (Meyer et al., 2020; Larsen et al., 2021; Chun et al., 2021; Rongcai et al., 2023; Y. Wang et al., 2024; Xie et al., 2024), or based on other RL agents (Zhao and Roh, 2019; Cui et al., 2023; Guan et al., 2024). In this article, we train an RL agent using PPO within an environment that includes a rule-compliant, i.e., reactive traffic participant.

### 2.4. Scenario generation for maritime navigation

Beyond learning controllers, several studies use RL to adversarially control surrounding ships and generate challenging scenarios to evaluate autonomous maritime decision-making systems (Vatle et al., 2024; Zhu et al., 2025; Laursen and Mengshoel, 2025). Additionally, generating challenging and representative scenarios to test maritime motion planners has been explored through approaches such as manual specification of scenario sets (Sawada et al., 2024), discriminative neural networks (Porres et al., 2020), and formulating scenario synthesis as a constraint satisfaction problem (Frey et al., 2025). Automatic identification system (AIS) data have also been used to extract and characterize real-world traffic (Bakdi et al., 2021; W. Wang et al., 2024; Xin et al., 2025), generate synthetic scenarios from AIS-derived statistics (Zhu et al., 2022), and bias scenarios toward high-risk encounters (Shi et al., 2023). Robustness-guided falsification for autonomous vessels using the cross-entropy method with Gaussian mixture distribution sampling is proposed in He et al. (2022). As the robustness score, the sum of distances towards a collision over time is minimized. For the task of open-sea navigation, more nuanced robustness measures can be constructed from temporal logic formalizations of maritime traffic rules as proposed by Krasowski and Althoff (2021), Torben et al. (2023). Torben et al. (2023) apply falsification using Gaussian processes on vessel controllers to test their traffic rule compliance. While Torben et al. (2023) focus on testing existing controllers, our work integrates falsification into the RL training process. To quantitatively evaluate the rule-compliance of scenarios, we extend the formalization by Krasowski and Althoff (2021) with robustness measures.

## 3. Preliminaries and problem statement

To formulate our method, we first present the vessel dynamics and relevant notation. We then outline STL, the formal language used to specify falsification targets. Finally, we present the formalized maritime traffic rules and state the problem statement of this work.

### 3.1. Notation and system dynamics

The Minkowski sum of a scalar $a \in \mathbb{R}$ and a set $B \subseteq \mathbb{R}$ is defined as $a \oplus B := \{a + b \mid b \in B\}$. The sign function $\mathrm{sgn}(a)$ equals 1 if $a > 0$ and $-1$ otherwise. The 2D rotation matrix for a counterclockwise rotation by an angle $\beta \in \mathbb{R}$ is returned by $\mathrm{R}(\beta)$. The state of a vessel $x \in \mathbb{R}^5$ is defined by the position $p = [p_x, p_y]^\top \in \mathbb{R}^2$ in the Cartesian coordinate frame, the orientation $\theta \in \mathbb{R}$, the velocity $v \in \mathbb{R}$ in the direction of $\theta$, and the angular velocity $\omega \in \mathbb{R}$. The vessel dynamics is:

$$\frac{d}{dt}\begin{bmatrix} p_x \\ p_y \\ \theta \\ v \\ \omega \end{bmatrix} = \begin{bmatrix} \cos(\theta)v \\ \sin(\theta)v \\ \omega \\ a \\ \alpha \end{bmatrix} = f(x, u), \tag{1}$$

where the input $u = [a, \alpha]^\top$ consists of the acceleration $a \in \mathbb{R}$ in the direction of $\theta$ and the angular acceleration $\alpha \in \mathbb{R}$. The velocities and accelerations are bounded by $v_{\min} \le v \le v_{\max}$, $|\omega| \le \omega_{\max}$, $|a| \le a_{\max}$, and $|\alpha| \le \alpha_{\max}$. We discretize the continuous vessel dynamics in (1) with respect to time using the Tsitouras 5(4) Runge-Kutta method (Tsitouras, 2011). Let $u_k$ and $x_k$ denote the input and state at time step $k \in \mathbb{Z}_{\ge 0}$. Each time step corresponds to the continuous time $k\Delta t$, with $\Delta t \in \mathbb{R}$ as the fixed time increment. The resulting discrete-time system $x_k = \Phi(x_{k-1}, u_{k-1})$ for $k \ge 1$ maps an initial state $x_0$ and an input sequence $\mathbf{u} := (u_0, u_1, \dots, u_{N-1})$ to a state sequence $\mathbf{x} := (x_1, x_2, \dots, x_N)$. We denote sequences in boldface to distinguish them from individual states and inputs. The value of any variable $\square$ at step $k$ associated with a vessel E is denoted as $\square_k^{\mathrm{E}}$. The joint state sequence is $\boldsymbol{\eta} = (\eta_0, \eta_1, \dots, \eta_N)$, where $\eta_k$ contains the states of all vessels at step $k$.

### 3.2. Signal temporal logic

The syntax of an STL formula $\varphi$ is (Bartocci et al., 2018, Sec. 2.1):

$$\varphi := \mu \mid \neg\varphi \mid \varphi_1 \vee \varphi_2 \mid \varphi_1 \mathbf{U}_{\mathrm{I}} \varphi_2 \,, \tag{2}$$

where $\mu := h_\mu(\eta_k) > 0$ is an atomic predicate defined by the robustness measure function $h_\mu : \mathbb{R}^n \to \mathbb{R}$, $\neg$ and $\vee$ are the Boolean operators for negation and disjunction, and $\mathbf{U}_{\mathrm{I}}$ is the until operator, where I is of the form $[a, b]$ or $[a, \infty)$, with $a, b \in \mathbb{Z}_{\ge 0}$ and $a \le b$. We denote that a state sequence $\boldsymbol{\eta}$ satisfies $\varphi$ at time step $k$ as $(\boldsymbol{\eta}, k) \models \varphi$ and write $\boldsymbol{\eta} \models \varphi$ when $k = 0$. The STL formula $(\boldsymbol{\eta}, k) \models \varphi_1 \mathbf{U}_{\mathrm{I}} \varphi_2$ holds if $\varphi_1$ holds until the formula $\varphi_2$ holds within $k \oplus \mathrm{I}$. Additional constants and operators can be derived from (2), e.g., the true constant $\top := \varphi \vee \neg\varphi$, the implication connective $\varphi_1 \implies \varphi_2 := \neg\varphi_1 \vee \varphi_2$, the eventually operator $\mathbf{F}_{\mathrm{I}}\varphi := \top \mathbf{U}_{\mathrm{I}} \varphi$, and the always operator $\mathbf{G}_{\mathrm{I}}\varphi := \neg \mathbf{F}_{\mathrm{I}} \neg\varphi$. The formula $(\boldsymbol{\eta}, k) \models \mathbf{F}_{\mathrm{I}}\varphi$ holds if there exists a time step in $k \oplus \mathrm{I}$ where $\varphi$ holds. Conversely, $(\boldsymbol{\eta}, k) \models \mathbf{G}_{\mathrm{I}}\varphi$ holds if $\varphi$ is true at every time step within $k \oplus \mathrm{I}$ (Bartocci et al., 2018, Sec. 2.1).

The robustness of complying with an STL formula quantifies the margin towards its violation. The robustness measure $\rho(\varphi, \boldsymbol{\eta}, k)$ of $\boldsymbol{\eta}$ with respect to an STL formula $\varphi$ at time step $k$ satisfies $\rho(\varphi, \boldsymbol{\eta}, k) > 0 \Leftrightarrow (\boldsymbol{\eta}, k) \models \varphi$ and is defined as (Donzé and Maler, 2010, Def. 1):

$$\begin{aligned}
\rho(\mu, \boldsymbol{\eta}, k) &:= h_\mu(\eta_k)\,, \\
\rho(\neg\varphi, \boldsymbol{\eta}, k) &:= -\rho(\varphi, \boldsymbol{\eta}, k)\,, \\
\rho(\varphi_1 \vee \varphi_2, \boldsymbol{\eta}, k) &:= \max\left(\rho(\varphi_1, \boldsymbol{\eta}, k), \rho(\varphi_2, \boldsymbol{\eta}, k)\right)\,, \\
\rho(\varphi_1 \mathbf{U}_{\mathrm{I}} \varphi_2, \boldsymbol{\eta}, k) &:= \max_{i \in k \oplus \mathrm{I}} \left( \min \left( \rho(\varphi_2, \boldsymbol{\eta}, i), \min_{j \in [k, i)} \rho(\varphi_1, \boldsymbol{\eta}, j) \right) \right)\,.
\end{aligned} \tag{3}$$

This makes it possible to frame rule compliance testing as an optimization problem that minimizes the robustness $\rho(\varphi, \boldsymbol{\eta}) := \rho(\varphi, \boldsymbol{\eta}, 0)$. While we use STL in this article, other temporal logics with quantitative semantics would also be applicable, e.g., metric temporal logic (Fainekos and Pappas, 2006).

### 3.3. Traffic rules

The Convention on the International Regulations for Preventing Collisions at Sea (COLREGs) (International Maritime Organization, 1972) specifies the traffic rules for open-sea navigation and defines three types of encounters between two power-driven vessels that potentially collide in the near future: crossing, head-on, and overtaking. In these encounters, there are two types of vessels: give-way and stand-on vessels (see Fig. 1). The give-way vessel has to adjust its course detectably, while the stand-on vessel has to maintain course and velocity until the situation is resolved. As specifications for the falsification, we adapt the formalization of maritime traffic rules from Krasowski and Althoff (2024). In particular, we focus on the three traffic rules where the autonomous vessel is the give-way vessel. These three rules are all instances of the specification $\varphi_{\texttt{colregs}}$:

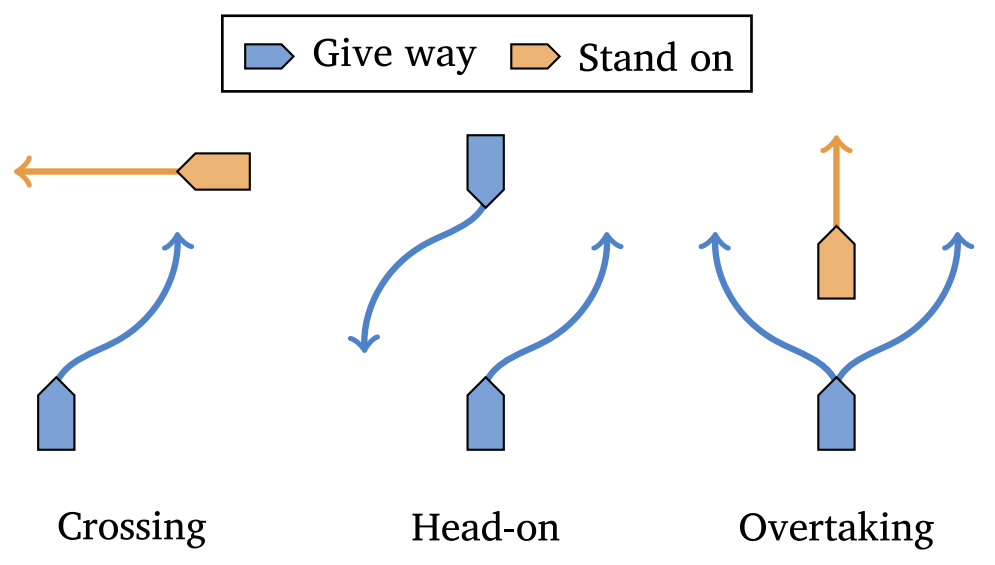

**Fig. 1.** Vessel encounters and rule-compliant maneuvers.

$$\begin{aligned}
\varphi_{\texttt{colregs}} :=\, &\mathbf{G}_{[0,\Delta t N]} \Big( \texttt{persistent_}\langle\texttt{encounter}\rangle(\mathbf{x}^{\mathrm{E}}, \mathbf{x}^{\mathrm{A}}, k) \implies \\
&\mathbf{F}_{[t_{\mathrm{p}}, t_{\mathrm{p}}+t_{\mathrm{m}}]} \left(\texttt{maneuver_}\langle\texttt{encounter}\rangle(\mathbf{x}^{\mathrm{E}}, k)\right) \wedge \\
&\mathbf{F}_{[t_{\mathrm{p}}, t_{\mathrm{p}}+2t_{\mathrm{m}}]} \left(\neg\texttt{velocity_obstacle}\left(\mathbf{x}^{\mathrm{E}}, \mathbf{x}^{\mathrm{A}}, k\right)\right) \Big)\,,
\end{aligned} \tag{4}$$

where ⟨encounter⟩ can take the values crossing, head_on, and overtake, and a persistent encounter persistent_⟨encounter⟩ is present if ⟨encounter⟩ is true for at least a duration of $t_{\mathrm{p}}$. For persistent encounters, the predicate maneuver_⟨encounter⟩ evaluates whether the required course adjustment is satisfied. The give-way maneuvers must be carried out within the time $t_{\mathrm{m}}$ in order to resolve the collision risk within $2t_{\mathrm{m}}$. For detailed derivations of the traffic rules, we refer the interested reader to Krasowski and Althoff (2021, 2024). We remark that the proposed approach can be applied to other rule sets as well.

### 3.4. Problem statement

The considered problem is learning an RL controller for maritime motion planning that satisfies traffic interaction rules two power-driven vessels governed by the COLREGs as defined in (4). With RL, the training scenario distribution dictates which behaviors a controller can learn. The unstructured nature of open-sea environments makes it challenging to identify traffic situations in which an RL policy violates the COLREGs. To address this, we use falsification-generated training scenarios that systematically expose the controller to rule-violating encounters. The focus of this article is on evaluating and demonstrating the improvements achieved by incorporating falsification into RL.

## 4. Falsification-driven reinforcement learning

For a given environment and safety specification, we aim to synthesize relevant training scenarios that define the initial conditions and the behavior of other dynamic entities. To this end, we use falsification to identify scenarios in which the current RL policy violates the specification and incorporate them into the training process (see Fig. 2).

Specifications with an implication structure, such as (4), can be satisfied in two ways. If the antecedent does not hold, the specification is vacuously satisfied. Otherwise, the consequent must hold, and the specification is nonvacuously satisfied. The robustness of STL implications is defined as the maximum of the negated robustness of the antecedent and the robustness of the consequent (see Section 3.2). As a result, it does not provide separate information about the individual robustness values of each component. Interface-aware STL (IA-STL) (Ferrère et al., 2019) addresses this issue by evaluating the robustness of the antecedent and the consequent separately: Let $\mathcal{I}$ and $\mathcal{O}$ denote the sets of atomic predicates in the antecedent and consequent, respectively. The IA-STL robustness, denoted as $\rho_{\mathcal{I}}^{\mathcal{O}}(\mu, \boldsymbol{\eta}, k)$, is analogous to (3), except for atomic predicates $\mu$ (Ferrère et al., 2019, Sec. 3):

$$\rho_{\mathcal{I}}^{\mathcal{O}}(\mu, \boldsymbol{\eta}, k) := \begin{cases} \rho(\mu, \boldsymbol{\eta}, k)\,, & \text{if } \mu \in \mathcal{I} \wedge \mu \notin \mathcal{O}\,, \\ \infty \cdot \operatorname{sgn}\left(\rho(\mu, \boldsymbol{\eta}, k)\right)\,, & \text{if } \mu \notin \mathcal{I} \wedge \mu \in \mathcal{O}\,, \\ 0\,, & \text{otherwise}\,. \end{cases} \tag{5}$$

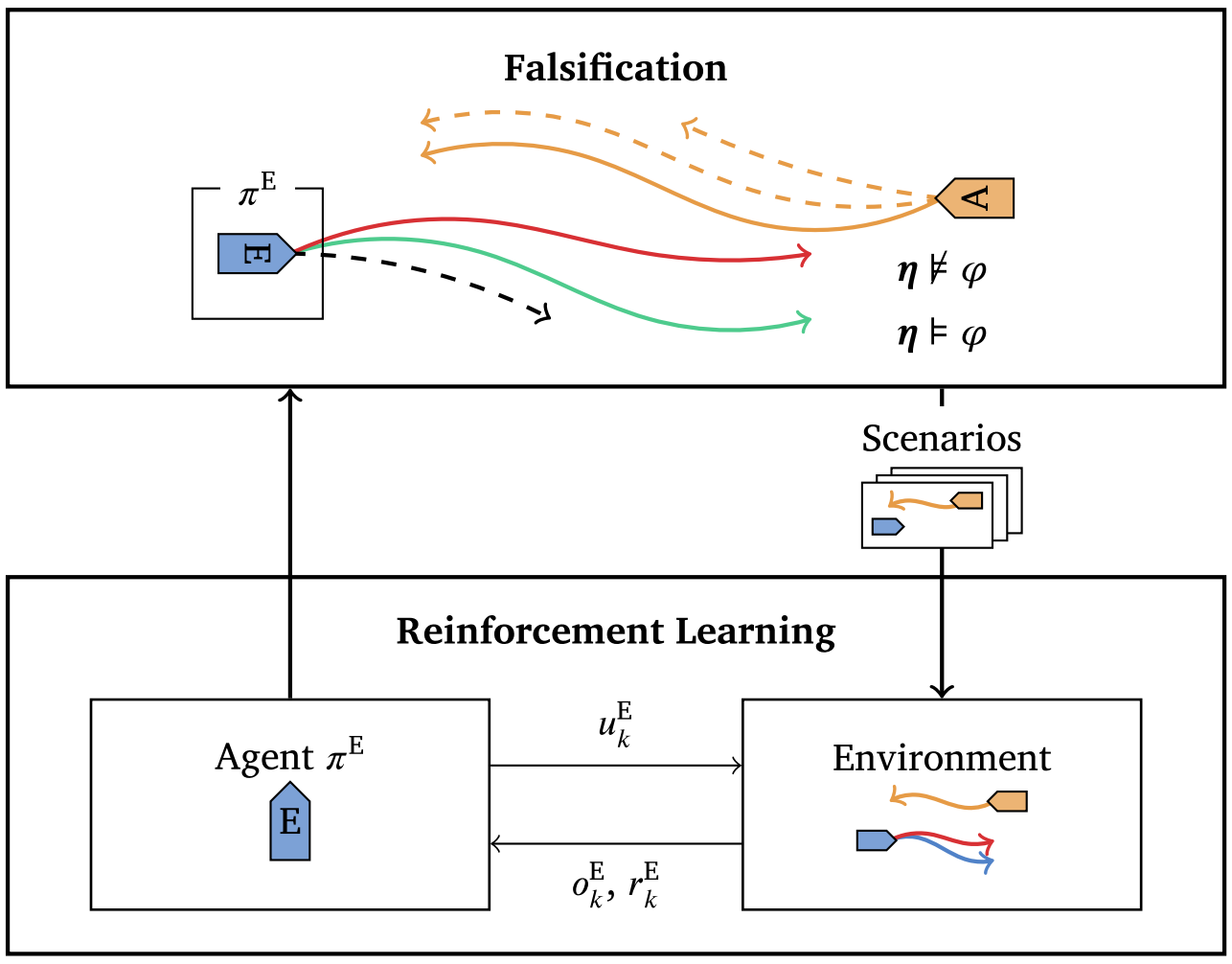

**Fig. 2.** Overview of the proposed framework. Every training iteration, the falsification process identifies scenarios where the agent policy $\pi^{\mathrm{E}}$ fails to comply with the specification $\varphi$. These scenarios are used for training the agent with standard reinforcement learning: the reward is $r_k^{\mathrm{E}}$, the observation is $o_k^{\mathrm{E}}$, and the agent action is $u_k^{\mathrm{E}}$.

If the antecedent does not hold, the input robustness $\rho_{\text{in}} := \rho_{\mathcal{I}}^{\emptyset}(\varphi, \boldsymbol{\eta})$ is positive and quantifies how far a scenario is from satisfying the antecedent. The output robustness $\rho_{\text{out}} := \rho_{\mathcal{O}}^{\mathcal{P}\setminus\mathcal{O}}(\varphi, \boldsymbol{\eta})$ is $\pm\infty$ in the vacuous case. Otherwise, it is finite and measures the robustness of the consequent. Fig. 3 illustrates vacuous and nonvacuous encounters (a) and the corresponding IA-STL input and output robustness values (b).

To generate scenarios that reveal specification violations, we solve the optimization problem in (6). We define a binary variable $\mathbf{1}_{\text{vacuous}}$ that indicates the vacuous case, i.e., $\mathbf{1}_{\text{vacuous}} := 1$ if $\rho_{\text{in}} > 0$, and $\mathbf{1}_{\text{vacuous}} := 0$ otherwise. In the vacuous case, the input robustness is minimized to achieve nonvacuity. Otherwise, the output robustness is minimized to falsify the specification. To prioritize nonvacuous scenarios, we use the Big-M method and add a large number $M$ to the objective in the vacuous case. The input sequence of the adversary $\mathbf{u}^{\mathrm{A}}$ is the decision variable, while the inputs of the ego system are computed based on the current policy $\pi^{\mathrm{E}}$. Moreover, to ensure that specification violations arise from the RL policy, the adversary must comply with its specifications $\varphi^{\mathrm{A}}$, resulting in:

$$
\begin{aligned}
\underset{\mathbf{u}^{\mathrm{A}}}{\arg\min} \quad & \mathbf{1}_{\text{vacuous}}(\rho_{\text{in}} + M) + (1 - \mathbf{1}_{\text{vacuous}})\rho_{\text{out}} \,, && (6)\\
\text{s.t.} \quad & x_k^{\mathrm{E}} = \boldsymbol{\Phi}\big(x_{k-1}^{\mathrm{E}}, \pi^{\mathrm{E}}(o_{k-1}^{\mathrm{E}})\big) \,, && k = 1, \dots, N \,,\\
& x_k^{\mathrm{A}} = \boldsymbol{\Phi}\big(x_{k-1}^{\mathrm{A}}, u_{k-1}^{\mathrm{A}}\big) \,, && k = 1, \dots, N \,,\\
& \mathbf{x}^{\mathrm{A}} \vDash \varphi^{\mathrm{A}} \,.
\end{aligned}
$$

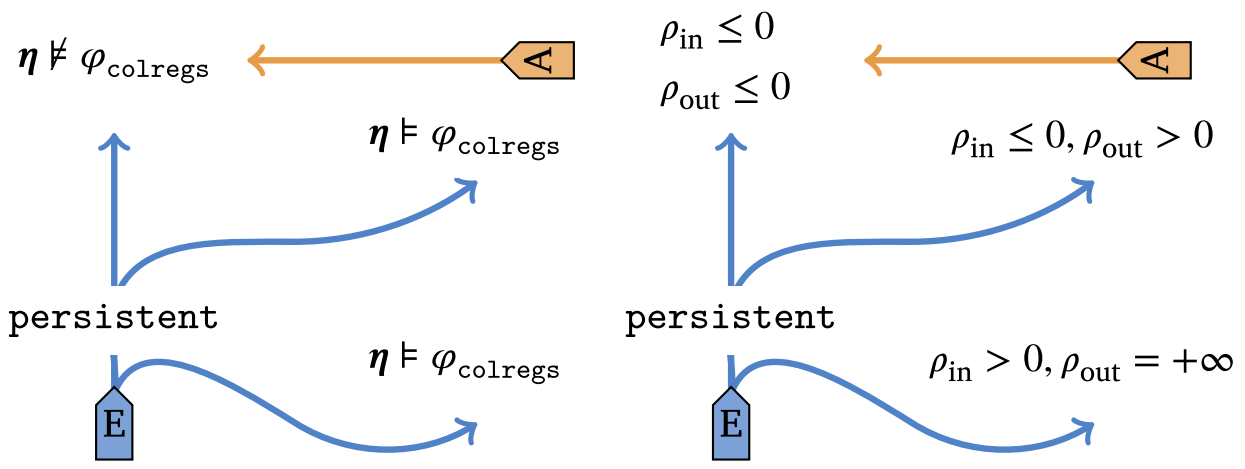

(a) Rule satisfaction (b) IA-STL robustness

**Fig. 3.** Illustration of (a) rule satisfaction and violation in a crossing encounter and (b) the corresponding IA-STL input and output robustness values.

Algorithm 1 presents our falsification-driven RL approach that is run for $T$ training steps. Every training iteration, we generate $n_{\text{samples}}$ new scenarios $(\chi_0, \mathbf{u}^{\mathrm{A}})$, where $\chi_0$ is the initialization of the scenario, and use CMA-ES, PSO, or SA to optimize for falsifying input sequences $\mathbf{u}^{\mathrm{A}}$ according to (6). The number of objective evaluations is limited to $n_{\text{fev}}$, if no falsifying scenario is found before. New scenarios are added to the training scenario pool $\mathcal{S}$ of size $n_{\text{pool}}$ that follows a first-in, first-out (FIFO) replacement strategy. The RL policy is trained on an environment that randomly samples scenarios from $\mathcal{S}$.

**Algorithm 1** Falsification-driven RL

**Input:** $T$, $\varphi$, $n_{\text{samples}}$, $n_{\text{pool}}$, $n_{\text{fev}}$
$\mathcal{S} \leftarrow \emptyset$, $i \leftarrow 0$ ▷ Scenario pool $\mathcal{S}$ of size $n_{\text{pool}}$
**while** $i < T$ **do**
  **for** $j = 1, \dots, n_{\text{samples}}$ **do**
    $\chi_0 \leftarrow$ SampleEnvironmentSetup( )
    $\mathbf{u}^{\mathrm{A}} \leftarrow$ Falsify$(\chi_0, \pi^{\mathrm{E}}, \varphi, n_{\text{fev}})$
    $\mathcal{S} \leftarrow \mathcal{S} \cup \{(\chi_0, \mathbf{u}^{\mathrm{A}})\}$ ▷ FIFO replacement
  **end for**
  $\mathcal{D} \leftarrow$ Rollout$(\pi^{\mathrm{E}}, \mathcal{S})$; $i \leftarrow i + |\mathcal{D}|$ ▷ Transition tuples $\mathcal{D}$
  RL update $\pi^{\mathrm{E}}$ with $\mathcal{D}$
**end while**

## 5. Maritime navigation

Falsification-driven RL is relevant in maritime navigation, as the unstructured nature of open-sea environments makes it challenging to identify traffic situations in which a learning policy behaves unsafely. We focus on two-vessel scenarios, where an ego vessel E should comply with $\varphi_{\texttt{colregs}}$, in an environment that includes one adversarial vessel A. With more adversarial vessels, it is easy to construct situations where the ego vessel faces conflicting constraints and must violate at least one traffic rule, e.g., acting as a stand-on vessel in an overtaking encounter while also needing to give way in a crossing encounter.

When the predicates `crossing` or `overtake` hold, A is the stand-on vessel and must maintain its course and velocity, i.e., we set $a_k^{\mathrm{A}} = 0$ and $\alpha_k^{\mathrm{A}}$ such that $\omega^{\mathrm{A}} \to 0$. In a persistent `head_on` encounter, A is a give-way vessel. Thus, we set $\alpha_k^{\mathrm{A}}$ so that the vessel turns approximately $\delta$ to starboard within $t_{\mathrm{m}}$, maintains this course for $t_{\mathrm{m}}$, then returns to the original course within another $t_{\mathrm{m}}$. Subsequently, we detail the RL observation and action spaces, the termination condition, and define the robustness measures for the atomic predicates of the traffic rules.

### 5.1. Reinforcement learning

**Agent interface** The RL action space is defined by the control inputs $u_k^{\mathrm{E}}$. We define an eleven-dimensional observation $o_k^{\mathrm{E}}$, which includes information about the ego state, the adversary vessel A, and a goal centered at $p^{\mathrm{G}}$. The ego observations consist of the linear velocity $v_k^{\mathrm{E}}$ and the angular velocity $\omega_k^{\mathrm{E}}$. The adversary observations consist of the Euclidean distance to the adversary, the angle from the ego orientation to the adversary, the adversary linear velocity $v_k^{\mathrm{A}}$, and the orientation difference relative to the ego, i.e., $\theta_k^{\mathrm{A}} - \theta_k^{\mathrm{E}}$. The goal observations are the distance to the goal center and the relative orientation toward the goal. All angles are represented by their cosine and sine components. Each vessel has a circular protected zone of radius $d_{\text{zone}}$ that must not intersect with the zones of other vessels. An episode terminates when the goal is reached, i.e., $\|p^{\mathrm{G}} - p_k^{\mathrm{E}}\| \leq d_{\text{goal}}$, or when protected zones intersect, i.e., $\|p_k^{\mathrm{A}} - p_k^{\mathrm{E}}\| \leq 2d_{\text{zone}}$.

**Reward function** The reward function consists of four dense and three sparse components. The dense components reward progress toward the goal, and penalize control effort, misalignment with the orientation toward the goal, and deviations from the nominal velocity range $[v_{\text{low}}, v_{\text{high}}]$. Specifically, the component $r_{\Delta\text{goal}}$ is proportional to the change in Euclidean distance from the goal center:

$$r_{\Delta\text{goal}} := c_{\Delta\text{goal}}(\|p^{\text{G}} - p^{\text{E}}_{k-1}\| - \|p^{\text{G}} - p^{\text{E}}_{k}\|)\,, \tag{7}$$

where $c_{\Delta\text{goal}}$ is a scaling coefficient. To promote smooth control behavior, $r_{\text{control}}$ penalizes the squared magnitude of the control input scaled by $c_{\text{control}}$, i.e., $r_{\text{control}} := c_{\text{control}}\|u^{\text{E}}_k\|^2$. The component $r_{\angle\text{goal}}$ penalizes deviations from the relative orientation toward the goal when not in a persistent encounter, scaled by $c_{\angle\text{goal}}$:

$$r_{\angle\text{goal}} := c_{\angle\text{goal}}\left(1 - \cos(\alpha_{\text{goal}} - \theta^{\text{E}}_k)\right)\,, \tag{8}$$

where $\alpha_{\text{goal}}$ is the angle from the ego position to the goal. Since vessels navigating open waters typically operate within a nominal speed range, the component $r_{\text{v}}$ applies a penalty, weighted by the coefficient $c_{\text{v}}$, when the linear velocity falls outside the range $[v_{\text{low}}, v_{\text{high}}]$:

$$r_{\text{v}} := \begin{cases} c_{\text{v}}(v^{\text{E}}_k - v_{\text{high}})\,, & \text{if } v^{\text{E}}_k > v_{\text{high}}\,, \\ c_{\text{v}}(v_{\text{low}} - v^{\text{E}}_k)\,, & \text{if } v^{\text{E}}_k < v_{\text{low}}\,, \\ 0\,, & \text{otherwise}\,. \end{cases} \tag{9}$$

The sparse components account for COLREGs compliance, incentivize reaching the goal, and penalize intersections of protected zones. Note that evaluating the satisfaction of $\varphi_{\texttt{colregs}}$ requires the complete joint state sequence $\boldsymbol{\eta}$. Providing a reward only at the final time step $N$ significantly increases the challenge of learning rule compliance due to the delayed feedback. We therefore define $\tilde{\varphi}_{\texttt{colregs}}$ as (4) without the outer $\mathbf{G}_{[0,\Delta t N]}$ operator. This allows $\tilde{\varphi}_{\texttt{colregs}}$ to be evaluated repeatedly with a delay of $t_{\text{p}} + 2t_{\text{m}}$, as derived in Maler et al. (2007, Def. 2). The reward component $r_{\text{colregs}}$ returns $c_{\neg\text{colregs}}$ for a violation of $\tilde{\varphi}_{\texttt{colregs}}$. The component $r_{\text{goal}}$ returns $c_{\text{goal}}$ when the ego vessel reaches the goal. The final component $r_{\text{zone}}$ assigns the penalty $c_{\text{zone}}$ when the protected zones intersect. The complete reward function is the sum of all components:

$$r := r_{\Delta\text{goal}} + r_{\text{control}} + r_{\angle\text{goal}} + r_{\text{v}} + r_{\text{colregs}} + r_{\text{goal}} + r_{\text{zone}}\,. \tag{10}$$

### 5.2. Robustness measures

The robustness measure functions $h_\mu$ are defined for six atomic predicates, based on which the metric temporal logic traffic rule formalization in Krasowski and Althoff (2024) can be transferred to STL. To address that differently scaled atomic predicates can cause masking effects in the robustness calculations and reduce falsification performance (Zhang et al., 2019), the robustness values are rescaled to the time domain, e.g., by dividing distances by the maximum velocity. Intuitively, the time domain scaling approximates the time required for vessels to violate (or satisfy) the predicate under worst-case conditions and, in doing so, provides a structured way to scale the robustness values. We categorize the atomic predicates into position, orientation, and velocity predicates, and then describe their connection to the formalization in Krasowski and Althoff (2024).

**Position predicate** The `position_halfplane` predicate reasons about the relative positioning of vessels. It determines if $p^{\text{A}}_k$ lies in the half-plane to the left of the line that passes through $p^{\text{E}}_k$ and is oriented at $\theta^{\text{E}}_k + \beta$, with $\beta \in [-\pi, \pi)$ (see Fig. 4a). The robustness is defined as the signed distance from $p^{\text{A}}_k$ to the line, divided by $v_{\max}$:

$$h_{\texttt{position_halfplane}}(x^{\text{E}}_k, x^{\text{A}}_k, \beta) := \frac{1}{v_{\max}}\begin{bmatrix} -\sin(\theta^{\text{E}}_k + \beta) \\ \cos(\theta^{\text{E}}_k + \beta)\end{bmatrix}^{\top}(p^{\text{A}}_k - p^{\text{E}}_k)\,. \tag{11}$$

**Orientation predicates** To reason about the relative orientations, the predicate `orientation_halfplane` is used. For $\gamma \in [-\pi, \pi)$, it determines if the vessel A is oriented to the left or right of $\theta^{\text{E}}_k + \gamma$ (see Fig. 4b). The robustness is defined as the minimal signed angular difference between $\theta^{\text{A}}_k$ and either $\theta^{\text{E}}_k + \gamma$ or $\theta^{\text{E}}_k + \gamma + \pi$, divided by $\omega_{\max}$:

$$h_{\texttt{orientation_halfplane}}(x^{\text{E}}_k, x^{\text{A}}_k, \gamma) := \frac{1}{\omega_{\max}}\arcsin\left(\sin(\theta^{\text{A}}_k - (\theta^{\text{E}}_k + \gamma))\right)\,. \tag{12}$$

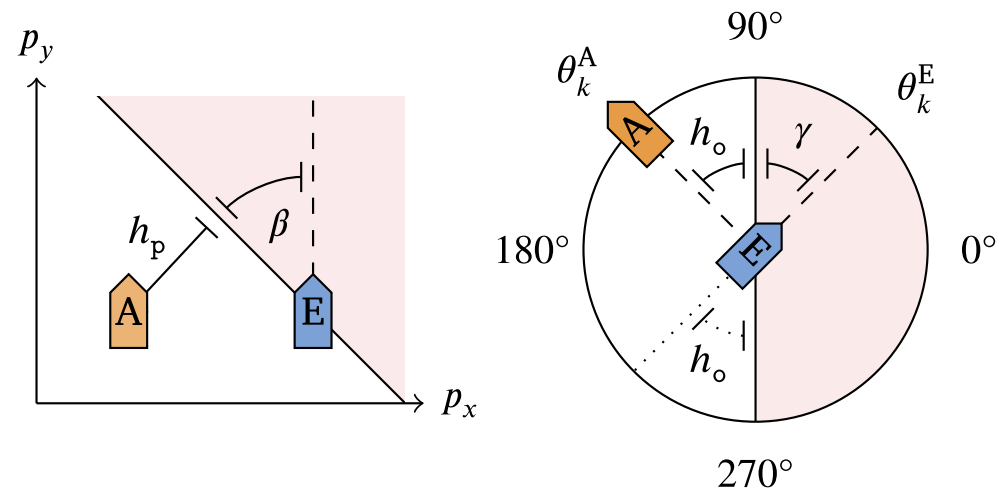

(a) $h_{\texttt{position_halfplane}}$　(b) $h_{\texttt{orientation_halfplane}}$

**Fig. 4.** Illustration of the robustness measures (a) $h_{\text{p}} := h_{\texttt{position_halfplane}}$ and (b) $h_{\text{o}} := h_{\texttt{orientation_halfplane}}$ with $\beta, \gamma := \pi/4$.

Note that arcsin and sin in (12) do not cancel each other, as $\theta^{\text{A}}_k - (\theta^{\text{E}}_k + \gamma)$ is not limited to $\left[-\frac{\pi}{2}, \frac{\pi}{2}\right]$. The give-way maneuvers are evaluated using the atomic predicate `change_course` to determine if the orientation change exceeds $\delta$. This predicate depends on the reference orientation $\theta_{\text{ref}}$, which is set to the orientation of the vessel performing the maneuver when a persistent encounter is detected. The robustness is defined as the signed angular difference between $\theta_{\text{ref}} + \delta$ and the current orientation of the vessel, divided by $\omega_{\max}$:

$$h_{\texttt{change_course}}(x^{\text{E}}_k, \delta) := \frac{1}{\omega_{\max}}(\theta^{\text{E}}_{\text{ref}} + \delta - \theta^{\text{E}}_k)\,. \tag{13}$$

**Velocity predicates** To detect a potential violation of the required separation between vessels, we define two tangent lines from $p^{\text{E}}_k$ to a circle of radius $2d_{\text{zone}}$ centered at $p^{\text{A}}_k$ to construct a velocity obstacle (Fiorini and Shiller, 1998). A velocity obstacle is the set of relative velocities that cause the protected zones to intersect if the vessels were to maintain their velocities and orientations from the time of evaluation onward. For details on its construction, we refer to Krasowski and Althoff (2021). Intuitively, we define the robustness via the required change in velocity to reach the boundary of the velocity obstacle. Let $\epsilon_k$ be the angle between the vector to the other vessel and the counterclockwise tangent line at time step $k$:

$$\epsilon_k := \arcsin\left(\frac{2d_{\text{zone}}}{\|p^{\text{A}}_k - p^{\text{E}}_k\|}\right)\,. \tag{14}$$

Furthermore, let $v^{\text{E,A}}_k$ be the relative velocity vector between vessel E and vessel A:

$$v^{\text{E,A}}_k := v^{\text{E}}_k\begin{bmatrix}\cos(\theta^{\text{E}}_k) \\ \sin(\theta^{\text{E}}_k)\end{bmatrix} - v^{\text{A}}_k\begin{bmatrix}\cos(\theta^{\text{A}}_k) \\ \sin(\theta^{\text{A}}_k)\end{bmatrix}\,. \tag{15}$$

The robustness of the predicate `velocity_halfplane` is the signed perpendicular distance (in velocity space) from $v^{\text{E,A}}$ to the tangent line, divided by $a_{\max}$:

$$h_{\texttt{velocity_halfplane}}(x^{\text{E}}_k, x^{\text{A}}_k, \epsilon_k) := \frac{1}{a_{\max}\|p^{\text{A}}_k - p^{\text{E}}_k\|}\left(\text{R}\left(\epsilon_k + \frac{\pi}{2}\right)(p^{\text{A}}_k - p^{\text{E}}_k)\right)^{\top} v^{\text{E,A}}\,. \tag{16}$$

For determining if the violation of the zone can happen within a bounded time horizon $t_{\text{h}}$, we define the robustness of the predicate `time_horizon`, which is the difference between $\|v^{\text{E,A}}\|$ and the minimum velocity required to travel the distance $\|p^{\text{A}}_k - p^{\text{E}}_k\|$, divided by $a_{\max}$:

$$h_{\texttt{time_horizon}}(x^{\text{E}}_k, x^{\text{A}}_k) := \frac{1}{a_{\max}}\left(\|v^{\text{E,A}}\| - \frac{\|p^{\text{A}}_k - p^{\text{E}}_k\|}{t_{\text{h}}}\right)\,. \tag{17}$$

The atomic predicate `drives_faster` determines whether the velocity of the ego vessel is greater than that of the other vessel in an overtaking encounter. The robustness measure is the difference in their velocities, divided by $a_{\max}$:

$$h_{\texttt{drives_faster}}(x_k^{\mathrm{E}}, x_k^{\mathrm{A}}) := \frac{1}{a_{\max}}(v_k^{\mathrm{E}} - v_k^{\mathrm{A}}) . \tag{18}$$

### 5.3. Formalized traffic rules

The atomic predicates are used to build the STL traffic rules $\varphi_{\texttt{colregs}}$, following the formalization in Krasowski and Althoff (2021, 2024). We consider three give-way traffic rules. They share the same high-level structure as defined in (4) and mainly differ in parameter values. For all give-way rules, an encounter predicate must persist for $t_{\mathrm{p}}$:

$$\begin{aligned}&\texttt{persistent_}\langle\texttt{encounter}\rangle(\mathbf{x}^{\mathrm{E}}, \mathbf{x}^{\mathrm{A}}, k) \iff \\ &\neg\langle\texttt{encounter}\rangle(\mathbf{x}^{\mathrm{E}}, \mathbf{x}^{\mathrm{A}}, k) \wedge \\ &\mathbf{G}_{[\Delta t, t_{\mathrm{p}}]}\left(\langle\texttt{encounter}\rangle(\mathbf{x}^{\mathrm{E}}, \mathbf{x}^{\mathrm{A}}, k)\right) ,\end{aligned} \tag{19}$$

The encounter predicates are $\{\texttt{crossing}, \texttt{overtake}, \texttt{head_on}\}$, which are all a conjunction of a relative position, a relative orientation, and a collision possibility predicate. We exemplify the three encounter predicates with the crossing encounter predicate:

$$\begin{aligned}&\texttt{crossing}(\mathbf{x}^{\mathrm{E}}, \mathbf{x}^{\mathrm{A}}, k, \underline{\beta}, \overline{\beta}, \underline{\gamma}, \overline{\gamma}) \iff \\ &\texttt{in_position}(\mathbf{x}^{\mathrm{E}}, \mathbf{x}^{\mathrm{A}}, k, \underline{\beta}, \overline{\beta}) \wedge \\ &\texttt{in_orientation}(\mathbf{x}^{\mathrm{E}}, \mathbf{x}^{\mathrm{A}}, k, \underline{\gamma}, \overline{\gamma}) \wedge \\ &\texttt{velocity_obstacle}(\mathbf{x}^{\mathrm{E}}, \mathbf{x}^{\mathrm{A}}, k) ,\end{aligned} \tag{20}$$

where $\underline{\beta}, \overline{\beta} \in [-\pi, \pi)$ and $\underline{\gamma}, \overline{\gamma} \in [-\pi, \pi)$ are angular bounds such that $\overline{\beta} > \underline{\beta}$, $\overline{\gamma} > \underline{\gamma}$, and $\overline{\beta} - \underline{\beta} \leq \pi$, $\overline{\gamma} - \underline{\gamma} \leq \pi$. The in_position predicate detects if $p_k^{\mathrm{A}}$ lies in the angular sector $[\theta_k^{\mathrm{E}} + \underline{\beta}, \theta_k^{\mathrm{E}} + \overline{\beta}]$, relative to $p_k^{\mathrm{E}}$:

$$\begin{aligned}&\texttt{in_position}(\mathbf{x}^{\mathrm{E}}, \mathbf{x}^{\mathrm{A}}, k, \underline{\beta}, \overline{\beta}) \iff \\ &\texttt{position_halfplane}(\mathbf{x}^{\mathrm{E}}, \mathbf{x}^{\mathrm{A}}, k, \underline{\beta}) \wedge \\ &\neg\texttt{position_halfplane}(\mathbf{x}^{\mathrm{E}}, \mathbf{x}^{\mathrm{A}}, k, \overline{\beta}) .\end{aligned} \tag{21}$$

Similarly, the in_orientation predicate detects whether $\theta_k^{\mathrm{A}}$ lies in the angular sector $[\theta_k^{\mathrm{E}} + \underline{\gamma}, \theta_k^{\mathrm{E}} + \overline{\gamma}]$:

$$\begin{aligned}&\texttt{in_orientation}(\mathbf{x}^{\mathrm{E}}, \mathbf{x}^{\mathrm{A}}, k, \underline{\gamma}, \overline{\gamma}) \iff \\ &\texttt{orientation_halfplane}(\mathbf{x}^{\mathrm{E}}, \mathbf{x}^{\mathrm{A}}, k, \underline{\gamma}) \wedge \\ &\neg\texttt{orientation_halfplane}(\mathbf{x}^{\mathrm{E}}, \mathbf{x}^{\mathrm{A}}, k, \overline{\gamma}) ,\end{aligned} \tag{22}$$

with position_halfplane and orientation_halfplane being defined in (11) and (12). The final predicate in (20) is velocity_obstacle, which evaluates whether the protected zones of the vessels intersect in the time horizon $t_{\mathrm{h}}$ given the vessels keep their course and speed. It is formally defined as:

$$\begin{aligned}&\texttt{velocity_obstacle}(\mathbf{x}^{\mathrm{E}}, \mathbf{x}^{\mathrm{A}}, k, \epsilon_k) \iff \\ &\texttt{velocity_halfplane}(\mathbf{x}^{\mathrm{E}}, \mathbf{x}^{\mathrm{A}}, k, -\epsilon_k) \wedge \\ &\neg\texttt{velocity_halfplane}(\mathbf{x}^{\mathrm{E}}, \mathbf{x}^{\mathrm{A}}, k, \epsilon_k) \wedge \\ &\texttt{time_horizon}(\mathbf{x}^{\mathrm{E}}, \mathbf{x}^{\mathrm{A}}, k) ,\end{aligned} \tag{23}$$

where $\epsilon_k$, velocity_halfplane and time_horizon are defined in (14), (16) and (17), respectively. The predicates overtake and head_on, and the values of $\beta$ and $\gamma$, are specified in Krasowski and Althoff (2021).

Having specified the encounter predicates, the remaining component in (4) to construct $\varphi_{\texttt{colregs}}$ is the maneuver predicate, which encodes the required course adjustment for the give-way vessel:

$$\begin{aligned}&\texttt{maneuver_\{crossing, head_on\}}(\mathbf{x}^{\mathrm{E}}, k, \delta) \iff \\ &\texttt{change_course}(\mathbf{x}^{\mathrm{E}}, k, -\delta) ,\end{aligned} \tag{24}$$

where change_course is defined in (13) and specifies the required course adjustment of $\delta$. For overtaking, the vessel has to turn to the left or right direction:

$$\begin{aligned}&\texttt{maneuver_overtake}(\mathbf{x}^{\mathrm{E}}, k, \delta) \iff \\ &\texttt{change_course}(\mathbf{x}^{\mathrm{E}}, k, -\delta) \vee \\ &\neg\texttt{change_course}(\mathbf{x}^{\mathrm{E}}, k, \delta) .\end{aligned} \tag{25}$$

## 6. Numerical evaluation

Next, we evaluate the falsification-driven RL framework using the RL formulation from Section 5.1 and the robustness measures of $\varphi_{\texttt{colregs}}$ from Section 5.2. We first describe the experimental setup, including the software packages, the parametrization, and the evaluation approach, and then report the results[2].

### 6.1. Experimental setup

We use Diffrax (Kidger, 2021) to discretize the vessel dynamics and evosax (Lange, 2023) to perform falsification. The policy and value networks share a two-layer multilayer perceptron with 128 units per layer and tanh activations, followed by a Long Short-Term Memory (LSTM) layer (Hochreiter and Schmidhuber, 1997) with 128 hidden units. The RL policy is trained with PPO using Flax (Heek et al., 2024) and Optax (Babuschkin et al., 2020), based on the implementation provided by Huang et al. (2022).

Table A.4 summarizes the parameters used in our numerical experiments. The traffic rule parameters are as proposed by Krasowski and Althoff (2024), which were set based on domain knowledge for open-sea navigation. For each scenario sample, the goal and initial state of the ego vessel are uniformly sampled from the distributions $\mathcal{P}$ and $\mathcal{X}_0^{\mathrm{E}}$. The initial state of vessel A is uniformly sampled from $\mathcal{X}_0^{\mathrm{A,C}}$, $\mathcal{X}_0^{\mathrm{A,H}}$, or $\mathcal{X}_0^{\mathrm{A,O}}$, which are designed to provide suitable initial conditions to induce crossing, head-on, and overtaking encounters, respectively. Since critical encounter scenarios are rare in maritime data, we generate random scenarios to train a baseline policy. The initial states of both vessels are initialized the same way as in falsification-driven training. To produce more realistic state sequences than straight-line motion, adversary control inputs are sampled from a mean-reverting stochastic process with zero mean, a standard deviation of 0.02, and a mean-reversion rate of 0.15. For the baseline, new scenarios are randomly sampled at each episode. The results for the final policies are reported across ten random seeds, and each RL policy is trained for $T$ steps. We train four agents using PPO: a baseline without falsification and three falsification-driven agents using CMA-ES, PSO, and SA, respectively.

We evaluate the performance of the baseline and falsification-driven agents using two scenario distributions. The baseline evaluation distribution consists of 10,000 randomly generated scenarios, where control inputs of the adversary are sampled from the same process that is used to train the baseline. The falsification evaluation distribution consists of 10,000 scenarios drawn from the training scenario pool, sampled uniformly across falsification-driven policies, random seeds, and training iterations every $2 \cdot 10^6$ steps. To obtain the strongest possible baseline agent, we tune the PPO hyperparameters in the baseline setting using Bayesian optimization over 90 runs. We select the setting that maximizes the rate of nonvacuous, rule-compliant scenarios across all encounter types. For the falsification-driven policies, we use the same PPO configuration (see Table A.5).

### 6.2. Results

We first analyze differences in training behavior between the baseline and falsification-driven policies on the baseline evaluation distribution. We then evaluate the final policies on both the baseline and falsification evaluation distributions. Finally, we support our claims on improved training and rollout performance using hypothesis tests.

[2] Code available at fdrl-maritime.github.io

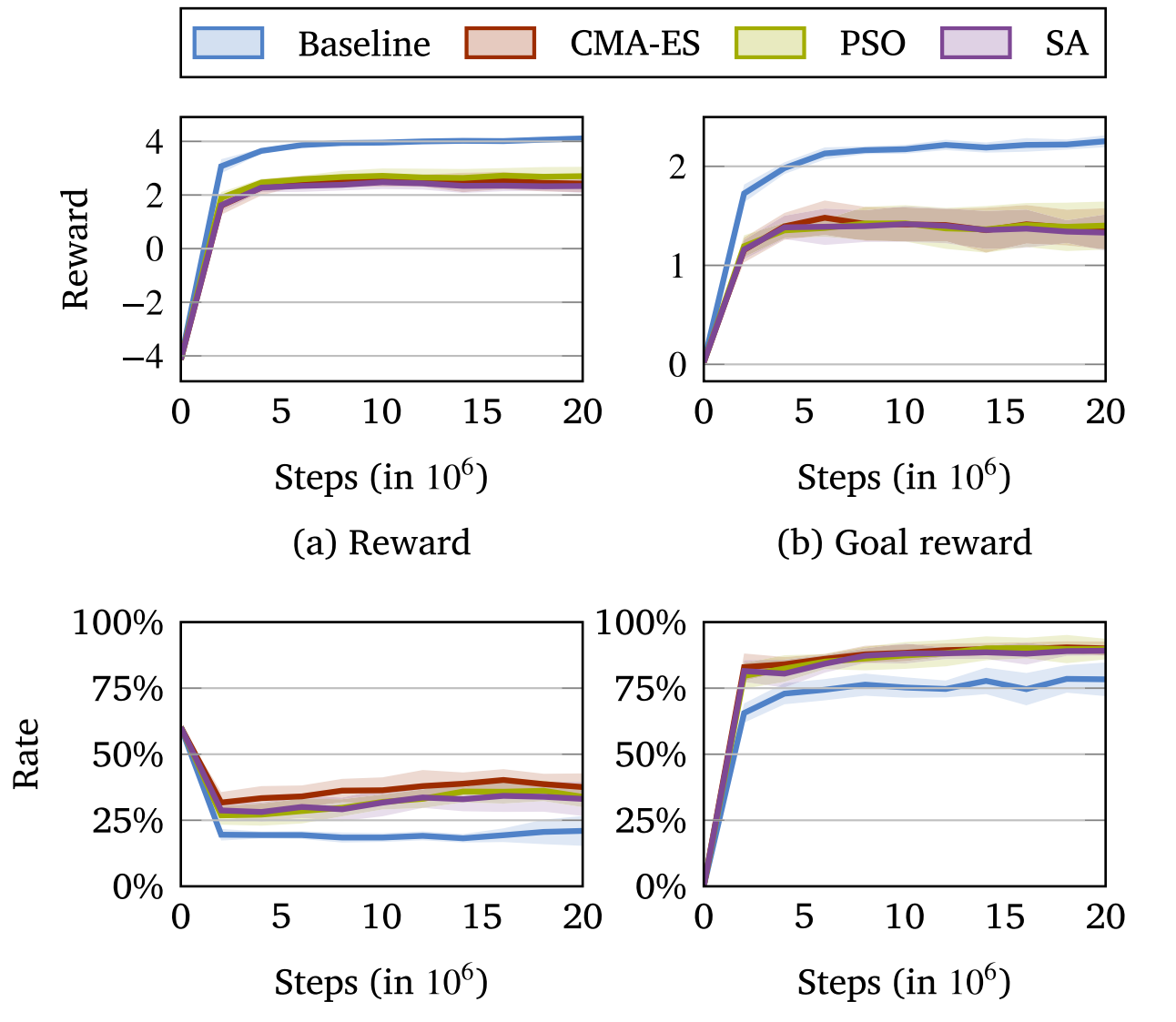

(a) Reward
(b) Goal reward
(c) Engagement
(d) Compliance

**Fig. 5.** Training behavior on the baseline evaluation scenarios. The mean and standard deviation (shaded) over random seeds are shown for (a) the RL reward, (b) the goal-reaching reward $r_{goal}$, (c) the ratio of nonvacuous encounters, and (d) nonvacuous rule compliance.

**Table 1**
Baseline distribution evaluation results (mean ± std). Compliance (†) includes vacuous scenarios; (‡) excludes them. All values are reported in percent (%) except robustness. The highest mean is shown in bold.

| Metric | Baseline | CMA-ES | PSO | SA |
|---|---|---|---|---|
| Goal | **75.1** ± 2.0 | 45.6 ± 6.9 | 46.7 ± 8.1 | 44.3 ± 6.1 |
| Compliance(†) | 95.7 ± 1.0 | 96.3 ± 1.1 | **96.6** ± 1.0 | 96.4 ± 0.9 |
| Robustness $\rho_{in}$ | **11.2** ± 2.7 | 6.0 ± 1.2 | 6.4 ± 0.8 | 7.4 ± 2.4 |
| Engagement | 21.0 ± 5.6 | **37.6** ± 5.2 | 33.9 ± 3.9 | 33.1 ± 6.4 |
| Compliance(‡) | 78.3 ± 6.3 | **90.0** ± 2.7 | 89.8 ± 3.9 | 89.2 ± 1.8 |
| Robustness $\rho_{out}$ | 10.9 ± 2.2 | 15.4 ± 3.0 | 15.2 ± 2.6 | **16.8** ± 3.8 |

During training, the baseline agent converges to a higher overall return with a notably higher goal-reaching reward compared to the falsification-driven policies (see Fig. 5). The falsification-driven policies maintain a higher proportion of nonvacuous scenarios, the so-called engagement, and converge to higher compliance rates than the baseline. Note that the baseline evaluation distribution used to compute the metrics in Fig. 5 differs from the distributions used to train the falsification-driven agents. Results on the falsification evaluation distribution over the course of training are provided in Appendix B.

Next to the training evaluation, we also investigate how the policies perform when testing them on the baseline evaluation distribution and the falsification evaluation distribution. The results on the baseline evaluation distribution (see Table 1) indicate the highest goal-reaching rate at 75.1% for the baseline policy. Compliance, including vacuous encounters, is above 95% for all methods. Compliance on nonvacuous encounters reduces to 78.3% for the baseline, and remains more stable for the falsification-driven policies with at least 89.2% and lower variance. The engagement is lower (i.e., 21%) for the baseline compared to 33.1% – 37.6% for the falsification-driven approaches, indicating that falsification-driven agents interact more often according to the COLREGs. Along the engagement, the input robustness is lower for the baseline agent than for the falsification-driven agents. The output robustness improves from 10.9 for the baseline to at least 15.2.

On the falsification evaluation distribution (see Table 2), the goal-reaching rate decreases for all methods to values between 26.3% and 30.6%. For the baseline policy, compliance, including vacuous encounters, decreases to 57.6%, with a negative mean input robustness and an engagement of 66.1%. Excluding vacuous encounters, the compliance of the baseline agent reduces further to 35.9%. The falsification-driven policies also show higher engagement rates between 47.1% and 49.8%, and maintain significantly higher compliance of at least 86.7% when including vacuous encounters and 71.7% when excluding them. The mean output robustness of the falsification-driven policies increases compared to the results on the baseline evaluation distribution. For the baseline policy, the output robustness even becomes negative (i.e., non-compliant).

**Table 2**
Falsification distribution evaluation results (mean ± std). Compliance (†) includes vacuous scenarios; (‡) excludes them. All values are reported in percent (%) except robustness. The highest mean is shown in bold.

| Metric | Baseline | CMA-ES | PSO | SA |
|---|---|---|---|---|
| Goal | **30.6** ± 2.0 | 27.4 ± 5.3 | 26.3 ± 5.8 | 27.0 ± 4.9 |
| Compliance(†) | 57.6 ± 2.1 | **86.8** ± 1.3 | 86.7 ± 1.8 | 86.7 ± 1.4 |
| Robustness $\rho_{in}$ | −7.2 ± 0.8 | −0.2 ± 1.4 | 0.3 ± 1.2 | **0.5** ± 1.8 |
| Engagement | **66.1** ± 2.5 | 49.8 ± 4.4 | 47.5 ± 2.7 | 47.1 ± 4.2 |
| Compliance(‡) | 35.9 ± 2.8 | **73.3** ± 3.3 | 72.0 ± 3.8 | 71.7 ± 1.4 |
| Robustness $\rho_{out}$ | −2.1 ± 1.9 | 19.2 ± 3.0 | 19.8 ± 2.5 | **20.7** ± 3.0 |

**Table 3**
Welch's t-test results for training and testing hypotheses with t-test statistic $t$ and degrees of freedom $df$. The p-value is $p < 0.001$ in all cases.

| | CMA-ES | | PSO | | SA | |
|---|---|---|---|---|---|---|
| **Metric** | $t$ | $df$ | $t$ | $df$ | $t$ | $df$ |
| Training | 51.9 | 17.9 | 51.8 | 16.8 | 44.5 | 17.3 |
| Testing | 3.62 | 16.4 | 3.84 | 17.4 | 4.09 | 14.5 |

Fig. 6 shows the relative frequency of input and output robustness values on the baseline evaluation distribution. The input robustness distribution of the baseline policy is shifted toward higher values compared to the falsification-driven policies, indicating less frequent engagement with the COLREGs. For the output robustness, we observe a peak around approximately −20, which corresponds to situations in which no maneuver is attempted and goal-reaching is prioritized at the risk of collision. This behavior occurs more frequently for the baseline policy than for the falsification-driven policies. When evaluated on the falsification evaluation distribution, the output robustness distributions of the falsification-driven policies exhibit a similar trend, while the baseline policy shows more pronounced safety-critical behavior (see Fig. B.9 in Appendix).

To provide intuition for the qualitative behavior of the learned policies, Fig. 7 presents representative rollouts of the trained baseline and a falsification-driven policy (CMA-ES) for crossing, head-on, and overtaking encounters. The rollouts use nine fixed initial positions, varying initial orientations, and set the angular acceleration of $\mathbf{u}^A$ to zero.

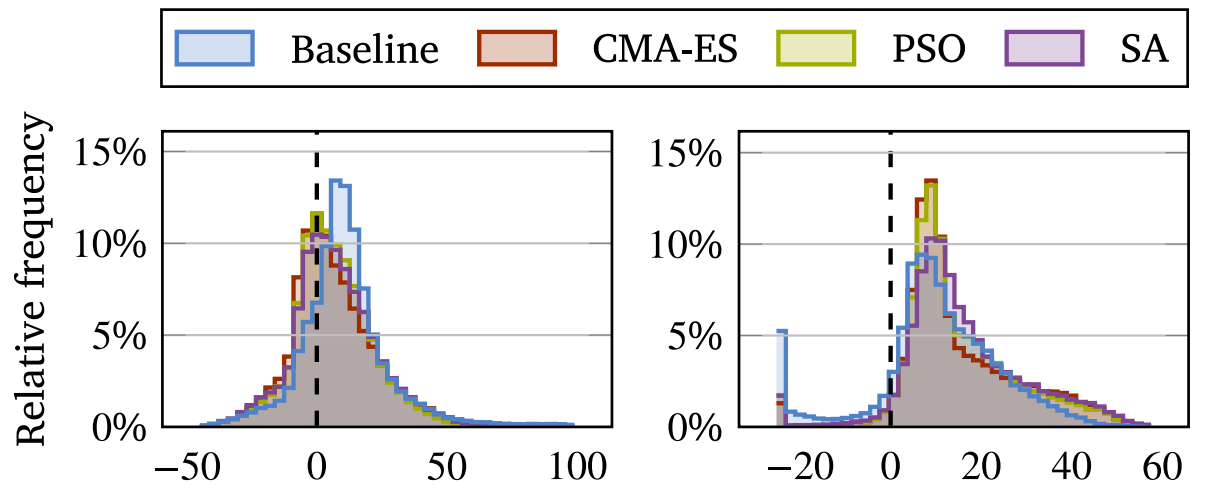

(a) Input robustness
(b) Output robustness

**Fig. 6.** Robustness distributions on the baseline evaluation scenarios.

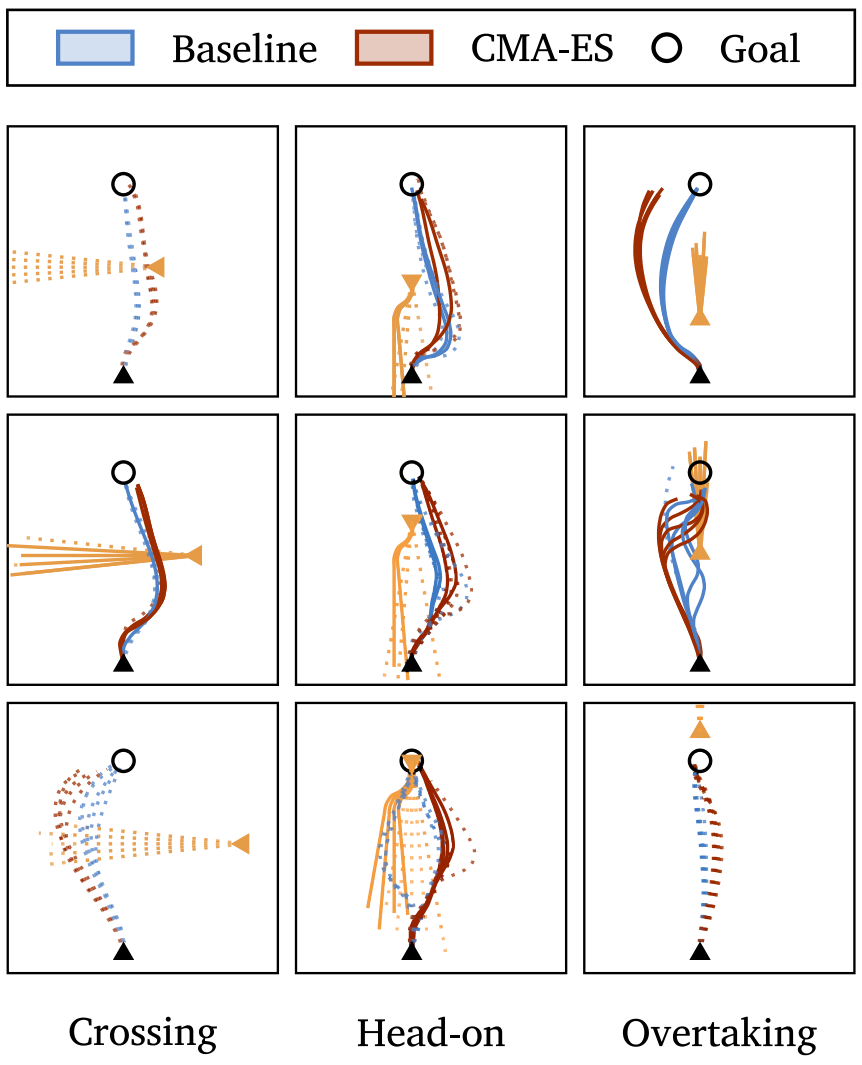

**Fig. 7.** Policy rollouts across encounter types. The adversary is shown in orange. Vacuous encounters are shown with dotted lines.

Particularly in the crossing and overtaking scenarios, the falsification-driven policy makes larger evasive maneuvers. This behavior illustrates the underlying trade-off between rule compliance and goal-reaching. In general, the agent should neither avoid encounters persistently nor deliberately induce them, as both strategies can compromise its ability to reach the goal. At the same time, prioritizing correct maneuvering over purely goal-directed behavior is necessary to ensure compliance with maritime traffic rules.

To statistically assess the central claims of this work, we conduct two hypothesis tests: one for training and one for testing. For training, we test whether falsification achieves smaller input robustness, requiring nonvacuous specification compliance and thus more relevant training scenarios (see Fig. 3). To this end, we evaluate the input robustness every $2 \cdot 10^6$ training steps for scenarios in the current falsification scenario pools or randomly sampled scenarios for baseline RL, respectively. For testing, we test whether falsification-trained policies achieve larger output robustness for nonvacuous scenarios on the baseline evaluation distribution, indicating more robust COLREGs compliance. Both tests are performed as one-sided, two-sample Welch's t-tests, which assume deviating variances of the alternative distributions. Each distribution provides for each seed (see Section 6.1) the mean of the relevant robustness value over all scenarios of that seed. Given small p-values and large t-test statistic and degree of freedom values (see Table 3), the numerical results indicate statistical significance for the test and training hypotheses.

## 7. Discussion

Our experiments indicate improved specification compliance compared to standard RL without falsification. To always guarantee compliance of the RL policy with the specifications, one can combine the learned policy with a safeguard for deployment (Krasowski et al., 2023). Specifically, two interesting directions for future research would be developing a capable provable fallback controller that can be invoked when the action proposed by the agent is not specification-compliant and extending the safeguard for discrete actions proposed by Krasowski and Althoff (2024) to a continuous action space. A challenge for these extensions is avoiding overly conservative solutions while proposing computationally efficient algorithms.

Our current setup focuses on clearly described traffic regulations between power-driven vessels that are satisfiable. Yet, our approach is generally applicable, and future work should focus on more complex traffic situations with more traffic participants and static obstacles. To apply falsification or verification methods in more complex settings, there must be extended formalized maritime traffic rules that are satisfiable. Naively applying the rules defined on two vessels to more vessels leads to traffic situations in which vessels need to violate some traffic rules. At the same time, showing that a set of rules is satisfiable in a continuous domain is challenging, even with simplified dynamics (Platzer and Clarke, 2009). Thus, future research should focus on developing such rules, which are a necessary precondition not only for falsification but also for assured interpretable behavior of autonomous vessels.

Achieving robust and performant policy behavior within environments governed by formal specifications, such as traffic regulations, requires training on a comprehensive array of rule-relevant scenarios. However, real-world datasets and nonadversarial generation techniques often fail to provide a sufficient ratio of these critical interactions. In addition, failure modes are commonly long-tail scenarios, which are increasingly rarely sampled from nonadversarial distributions (Ding et al., 2023). In contrast, adversarial scenario generation techniques achieve a higher rate of relevant scenarios that are tailored towards falsifying the current policy. As the computational cost per adversarial scenario is roughly constant, falsification-driven RL will become more efficient than training on nonadversarial distributions when aiming for, e.g., high rule compliance; see Theorem 1 in Liu and Feng (2024).

## 8. Conclusions

In this article, we introduce a novel falsification-driven RL approach. By integrating efficient, sampling-based falsification techniques into the RL training process, our approach leverages counterexamples to iteratively refine policy behavior and promote compliance with complex STL specifications. This article is the first to apply falsification-driven RL to maritime motion planning. Our experiments indicate that incorporating falsification improves the training process by generating more relevant scenarios, which in turn leads to more consistent rule compliance. These findings highlight the potential of combining formal methods with learning-based control to enhance the safety of autonomous systems. Future work should focus on scaling the approach to multi-agent scenarios and validating it in real-world maritime environments to bridge the gap between simulation and deployment.

## CRediT authorship contribution statement

**Marlon Müller:** Writing – review & editing, Writing – original draft, Software, Methodology, Conceptualization. **Florian Finkeldei:** Writing – review & editing, Writing – original draft, Software, Methodology, Conceptualization. **Hanna Krasowski:** Writing – review & editing, Writing – original draft, Software, Methodology, Conceptualization. **Murat Arcak:** Writing – review & editing, Supervision. **Matthias Althoff:** Writing – review & editing, Supervision.

## Declaration of competing interest

The authors declare that they have no known competing financial interests or personal relationships that could have appeared to influence the work reported in this paper.

## Acknowledgments

This work was funded by the Deutsche Forschungsgemeinschaft (DFG, German Research Foundation) through the grants SFB 1608, AL 1185/17-1, and AL 1185/19-1, by the National Science Foundation through the grant CNS-2111688, and by the Munich Center of Machine Learning (MCML). The authors gratefully acknowledge the scientific support and resources of the AI service infrastructure LRZ AI Systems provided by the Leibniz Supercomputing Centre (LRZ) of the Bavarian Academy of Sciences and Humanities (BAdW), funded by Bayerisches Staatsministerium für Wissenschaft und Kunst (StMWK).

## Appendix A. Parameters

Table A.4 and Table A.5 report the experimental parameters.

**Table A.4**
Experiment parameters.

| Parameter | Value | Parameter | Value |
|---|---|---|---|
| *System Dynamics* | | | |
| $\Delta t$ | 15 s | $N$ | 80 |
| $v_{\min}$ | $5\,\mathrm{m\,s^{-1}}$ | $v_{\max}$ | $15\,\mathrm{m\,s^{-1}}$ |
| $v_{\mathrm{low}}$ | $7.5\,\mathrm{m\,s^{-1}}$ | $v_{\mathrm{high}}$ | $10\,\mathrm{m\,s^{-1}}$ |
| $a_{\max}$ | $0.12\,\mathrm{m\,s^{-2}}$ | $\omega_{\max}$ | $0.015\,\mathrm{rad\,s^{-1}}$ |
| $\alpha_{\max}$ | $0.000\,25\,\mathrm{rad\,s^{-2}}$ | | |
| *Traffic Rules* | | | |
| $t_{\mathrm{m}}$ | 75 s | $d_{\mathrm{zone}}$ | 900 m |
| $t_{\mathrm{p}}$ | 60 s | $\delta$ | 20° |
| $t_{\mathrm{h}}$ | 420 s | | |
| *Initial State and Goal* | | | |
| $\mathcal{X}_0^{\mathrm{E}}$ | $\begin{pmatrix}[-1.5\,\mathrm{km}, 1.5\,\mathrm{km}]\\ [-6\,\mathrm{km}, -4\,\mathrm{km}]\\ [80^\circ, 100^\circ]\\ 7.5\,\mathrm{m\,s^{-1}}\\ 0^\circ\,\mathrm{s^{-1}}\end{pmatrix}$ | $\mathcal{X}_0^{\mathrm{A,C}}$ | $\begin{pmatrix}[4\,\mathrm{km}, 6\,\mathrm{km}]\\ [-2\,\mathrm{km}, 2\,\mathrm{km}]\\ [150^\circ, 210^\circ]\\ [2.5\,\mathrm{m\,s^{-1}}, 12.5\,\mathrm{m\,s^{-1}}]\\ 0^\circ\,\mathrm{s^{-1}}\end{pmatrix}$ |
| $\mathcal{X}^{\mathrm{A,H}}$ | $\begin{pmatrix}[-2\,\mathrm{km}, 2\,\mathrm{km}]\\ [4\,\mathrm{km}, 6\,\mathrm{km}]\\ [268^\circ, 272^\circ]\\ [5\,\mathrm{m\,s^{-1}}, 10\,\mathrm{m\,s^{-1}}]\\ 0^\circ\,\mathrm{s^{-1}}\end{pmatrix}$ | $\mathcal{X}_0^{\mathrm{A,O}}$ | $\begin{pmatrix}[-2\,\mathrm{km}, 2\,\mathrm{km}]\\ [-2\,\mathrm{km}, 1\,\mathrm{km}]\\ 90^\circ\\ [2.5\,\mathrm{m\,s^{-1}}, 4\,\mathrm{m\,s^{-1}}]\\ 0^\circ\,\mathrm{s^{-1}}\end{pmatrix}$ |
| $\mathcal{P}$ | $\begin{pmatrix}[-2\,\mathrm{km}, 2\,\mathrm{km}]\\ [4\,\mathrm{km}, 6\,\mathrm{km}]\end{pmatrix}$ | | |
| *Falsification* | | | |
| $n_{\mathrm{samples}}$ | 16 | $n_{\mathrm{fev}}$ | 320 |
| $n_{\mathrm{pool}}$ | 1024 | | |
| *Reinforcement Learning* | | | |
| $c_{\Delta\mathrm{goal}}$ | +0.0003 | $c_{\mathrm{v}}$ | −0.05 |
| $c_{\mathrm{control}}$ | −0.05 | $c_{\mathrm{goal}}$ | +3 |
| $c_{\neg\mathrm{colregs}}$ | −6 | $c_{\mathrm{zone}}$ | −3 |
| $c_{\angle\mathrm{goal}}$ | −0.1 | $d_{\mathrm{goal}}$ | 300 m |

**Table A.5**
PPO hyperparameters.

| Parameter | Value |
|---|---|
| Training steps $T$ | $20 \cdot 10^6$ |
| Learning rate | $5.314 \cdot 10^{-4}$ |
| Entropy coefficient | $0.00665 \to 0$ |
| Discount factor $\gamma$ | 0.99 |
| GAE $\lambda$ | 0.9 |
| Clip coefficient $\epsilon$ | 0.1 |
| Value function coefficient | 0.5 |
| Max gradient norm | 0.5 |
| Steps per rollout | 160 |
| Minibatch size | 5120 |
| Update epochs | 10 |
| Action distribution | Tanh Gaussian |

## Appendix B. Evaluation on the falsification distribution

Fig. B.8 depicts the training behavior of the baseline and falsification-driven policies on the falsification evaluation distribution, while Fig. B.9 shows the relative frequency of input and output robustness values when evaluating the trained policies.

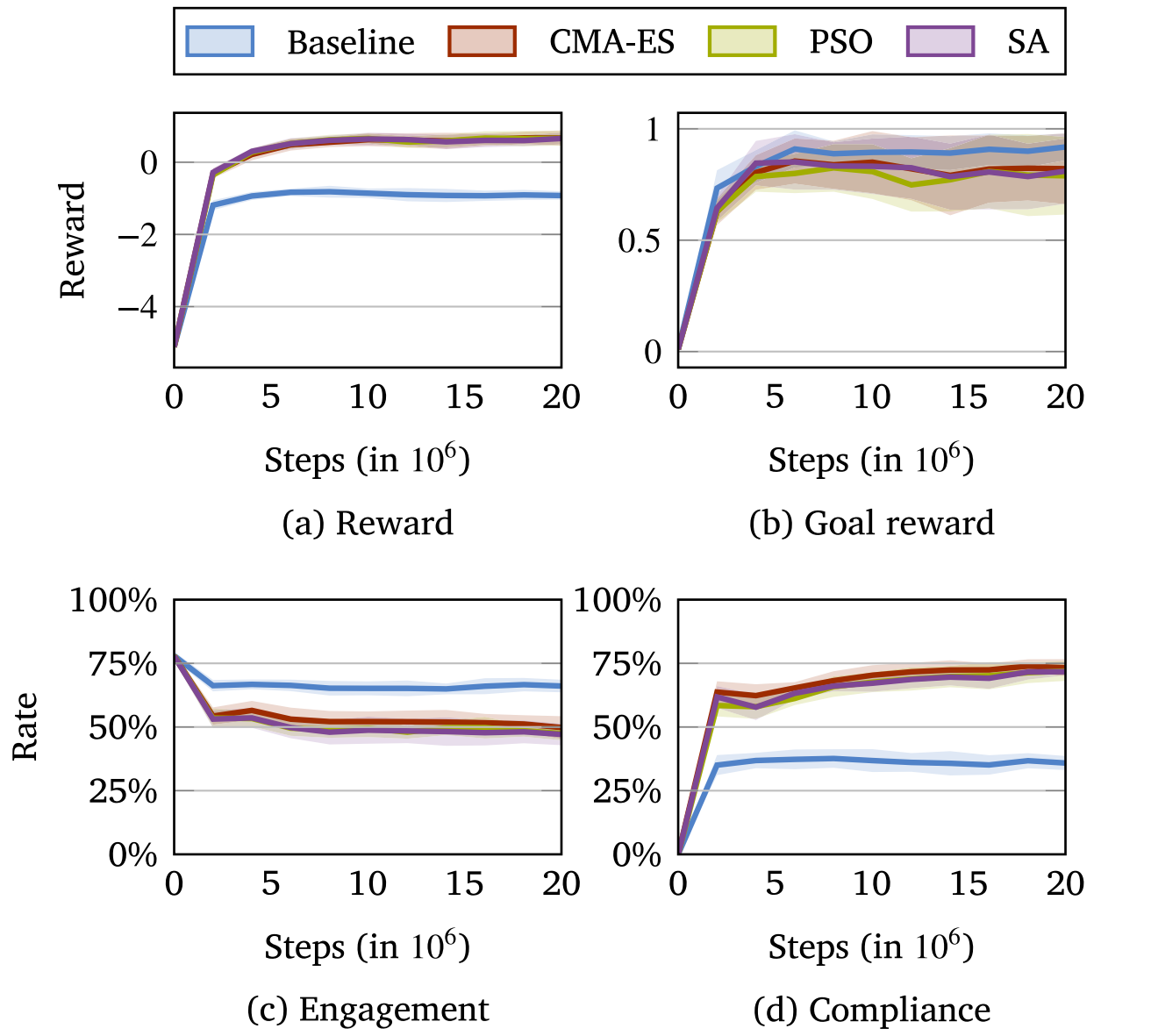

**Fig. B.8.** Training behavior on the falsification evaluation scenarios. The mean and standard deviation (shaded) over random seeds are shown for (a) the RL reward, (b) the goal-reaching reward $r_{\mathrm{goal}}$, (c) the ratio of nonvacuous encounters, and (d) nonvacuous rule compliance.

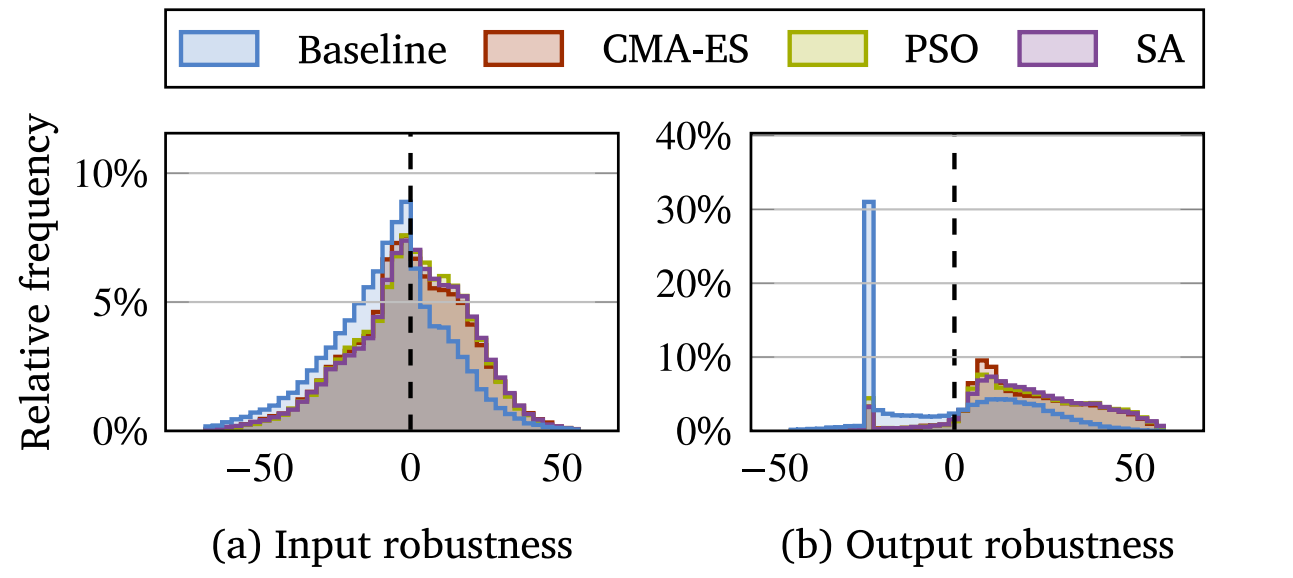

**Fig. B.9.** Robustness distributions on the falsification evaluation scenarios.